\documentclass[conference]{IEEEtran}
\IEEEoverridecommandlockouts
\usepackage{cite}
\usepackage{amsmath,amssymb,amsfonts}
\usepackage{algorithm,algorithmic}
\usepackage{graphicx}
\usepackage{textcomp}
\usepackage{xcolor}
\def\BibTeX{{\rm B\kern-.05em{\sc i\kern-.025em b}\kern-.08em
    T\kern-.1667em\lower.7ex\hbox{E}\kern-.125emX}}

\usepackage{url}
\usepackage{tcolorbox}
\usepackage{multirow}
\usepackage{diagbox}
\usepackage{arydshln}
\usepackage{comment}

\usepackage{hyperref}
\hypersetup{
    colorlinks=true,
    citecolor=blue,
    urlcolor=blue,
    }

\usepackage{inputenc}
\DeclareUnicodeCharacter{0308}{COMMAND}

\begin{document}

\title{Extracting Protein-Protein Interactions (PPIs) from Biomedical Literature using Attention-based Relational Context Information\\
\thanks{* Corresponding author.\\
The materials presented in this paper are based upon the work supported by the U.S. Department of Energy, Office of Science, Office of Biological and Environmental Research, DE-SC0012704.}
}

\makeatletter
\newcommand{\linebreakand}{%
  \end{@IEEEauthorhalign}
  \hfill\mbox{}\par
  \mbox{}\hfill\begin{@IEEEauthorhalign}
}
\makeatother

\author{
    \IEEEauthorblockN{Gilchan Park*}
    \IEEEauthorblockA{\textit{Computational Science Initiative} \\
        \textit{Brookhaven National Laboratory}\\
        Upton, New York, USA \\
        \tt{gpark@bnl.gov}}
    \and
    \IEEEauthorblockN{Sean McCorkle}
    \IEEEauthorblockA{\textit{Computational Science Initiative} \\
        \textit{Brookhaven National Laboratory}\\
        Upton, New York, USA \\
        \tt{mccorkle@bnl.gov}}
    \and
    \IEEEauthorblockN{Carlos Soto}
    \IEEEauthorblockA{\textit{Computational Science Initiative} \\
        \textit{Brookhaven National Laboratory}\\
        Upton, New York, USA \\
        \tt{csoto@bnl.gov}}
    \linebreakand 
    \IEEEauthorblockN{Ian Blaby}
    \IEEEauthorblockA{\textit{Joint Genome Institute} \\
        \textit{Lawrence Berkeley National Laboratory}\\
        Berkeley, California, USA \\
        \tt{ikblaby@lbl.gov}}
    \and
    \IEEEauthorblockN{Shinjae Yoo}
    \IEEEauthorblockA{\textit{Computational Science Initiative} \\
        \textit{Brookhaven National Laboratory}\\
        Upton, New York, USA \\
        \tt{sjyoo@bnl.gov}}
}

\maketitle

\begin{abstract}
Because protein-protein interactions (PPIs) are crucial to understand living systems, harvesting these data is essential to probe disease development and discern gene/protein functions and biological processes. 
Some curated datasets contain PPI data derived from the literature and other sources (e.g., IntAct, BioGrid, DIP, and HPRD). However, they are far from exhaustive, and their maintenance is a labor-intensive process.
On the other hand, machine learning methods to automate PPI knowledge extraction from the scientific literature have been limited by a shortage of appropriate annotated data.
This work presents a unified, multi-source PPI corpora with vetted interaction definitions augmented by binary interaction type labels and a Transformer-based deep learning method that exploits entities' relational context information for relation representation to improve relation classification performance.
The model's performance is evaluated on four widely studied biomedical relation extraction datasets, as well as this work's target PPI datasets, to observe the effectiveness of the representation to relation extraction tasks in various data. Results show the model outperforms prior state-of-the-art models. The code and data are available at:  \tt{\url{https://github.com/BNLNLP/PPI-Relation-Extraction}}
\end{abstract}


\section{Introduction}
Much effort in modern molecular biology either involves or is entirely focused on learning and understanding the functions and interactions of the millions of proteins that compose the basic building blocks of life. In particular, the prediction of protein structure and functions has been recognized as a paramount phase in some major issues of life science, such as the therapeutic approach for several diseases, which can ameliorate healthcare by accelerating drug discovery and development. 
The functions of most proteins currently are unknown with only a small fraction definitively established after extensive and labor-intensive lab work has been performed. 
These gold-standard protein function assignments have been extended computationally via DNA and amino acid sequence homology throughout the ever-expanding collection of protein sequences determined from genome sequencing.
However, inference from homology often is inaccurate. 
Helpfully, clues about function can come from other sources, including interactions with proteins for which the function is known.
While experiments that definitively determine interactions can be labor-intensive, several relatively high-throughput methods are in use, such as two-hybrid screening \cite{bruckner2009y2h} and affinity purification followed by mass spectrometry \cite{dunham2012affin}.
Numerous databases, such as IntAct\footnote{\tt{\url{https://www.ebi.ac.uk/intact}}}, STRING\footnote{\tt{\url{https://string-db.org}}}, DIP\footnote{\tt{\url{https://dip.doe-mbi.ucla.edu/dip}}}, BioGrid\footnote{\tt{\url{https://thebiogrid.org}}}, HPRD\footnote{\tt{\url{https://www.hprd.org}}}, and MINT\footnote{\tt{\url{https://mint.bio.uniroma2.it}}}
are now dedicated to collecting and curating protein-protein interaction (PPI) results obtained using various techniques and from the scientific literature.
Unfortunately, mining the literature requires manual effort and is slow.
To remedy this, we aim to develop a machine learning (ML) model that effectively identifies statements of PPIs in scientific text.

Efforts to fully automate text knowledge extraction are widespread and ongoing with supervised learning approaches currently being the most favored.
A key challenge in applying these methods to PPI extraction is a shortage of training data specifically annotated for this purpose.
Several publicly available PPI training datasets suffer from biases of restricted biological focus (i.e., human-, medical-, or microbial-only) and also differences in the concept of what \textit{defines} an interaction.
For this work, we combine all of the aforementioned training sets, vet them for uniformity in interaction definition, and add interaction type labels.
We also propose Transformer architecture-based models \cite{vaswani2017attention}, which leverage entities' relational context information to build a relation representation that improves relation classification performances.

As detailed in this paper, our contribution is twofold: 
\begin{enumerate}
\item We augment public PPI corpora with labels for protein types (\textit{enzyme} and \textit{structural}), which further delineate the functional role of proteins and consequently afford a helpful protein classification for the biology community. We also provide the interaction-typed PPI corpora for the community.

\item We present a Transformer-based relation prediction method that exploits entities' relational context information to build an improved relation representation. Our study shows the effectiveness of the proposed approach not only on the PPI datasets, but also four biomedical relation extraction datasets.
\end{enumerate}

\section{Related Work}

There have been ongoing efforts to consolidate biological knowledge pertinent to PPIs from literature by creating machine-processable data and designing protein relation extraction methods.

\subsection{PPI corpora}

BioCreative VI \cite{islamaj2019overview} proposed a PPI relation extraction challenge task related to genetic mutations to foster the development of mining PPI information from biomedical literature.
 Bunescu et al. \cite{bunescu2005comparative} annotated 1000 titles and abstracts from the MEDLINE repository that discuss human genes/proteins, the so-called AIMed corpus, which includes roughly 5000 protein names and 1000 protein interactions.
 Pyysalo et al. \cite{pyysalo2007bioinfer} created BioInfer (Bio Information Extraction Resource), containing 1100 sentences with named entities and their relationships tagged from abstracts of biomedical research articles.
Fundel, ̈K{\"u}ffner, and Zimmer \cite{fundel2007relex} tagged the sentences of 50 abstracts referenced by the Human Protein Reference Database (HPRD) with direct physical interactions, regulatory relations, and modifications between genes/proteins. 
The IEPA (Information Extraction Processing Assessment) corpus \cite{ding2001mining} was created to conduct a comparative study on the merits of different text processing units for interactions between biochemical entities. 
The Learning Language in Logic Workshop (LLL05) \cite{nedellec2005learning} designed the genic interaction extraction challenge task that aims to promote protein/gene interactions information extraction from biology abstracts in the MEDLINE bibliography database. 
The LLL challenge focused on gene interactions in \textit{Bacillus subtilis}, a model bacterium, and many papers have been published about direct gene interactions involved in sporulation.

Although the number of corpora and methods for PPI information extraction from biomedical text has increased as the interest in automatic mining systems has grown, the lack of consensus with respect to PPI annotation has hindered consolidation of heterogeneous datasets, thereby making it difficult for researchers to properly evaluate their methods on a standardized dataset for PPI extraction. Pyysalo et al. \cite{pyysalo2008comparative} have conducted a comparative analysis of the five PPI datasets---AIMed, BioInfer, HPRD50, IEPA, and LLL---and unified the PPI annotations to share with the community for clear and comparative method evaluation. 
To merge these diverse datasets, Pyysalo et al. \cite{pyysalo2008comparative} have found common categories across the five corpora and generated a unified PPI corpora composed of sentences tagged with undirected and untyped binary interactions (i.e., positive and negative). These unified versions of PPI datasets, hereafter called the \textit{five benchmark PPI corpora}, have been widely used to evaluate various approaches on PPI extraction tasks \cite{tikk2010comprehensive,bui2011hybrid,warikoo2021lbert}.
In the biological literature, single sentences often discuss more than two proteins, and such statements are not all declarations of interactions between the proteins mentioned.
These datasets include all identified protein/gene entity names found within each training sentence, as well as a pairwise evaluation of positive/negative interactions between each possible pairing.

However, some issues remain regarding the content and annotations in these benchmark PPI datasets (detailed in Section \ref{datase-problems}).
In this paper, we present an augmented, refined version of the five benchmark PPI corpora along with the BioCreative VI corpus that further specify positive interactions into two types of interactions: \textit{enzyme} and \textit{structural}. These interaction types are desirable to construct protein interaction networks.

\subsection{PPI extraction methods}

In the early stages of adopting ML approaches for the PPI extraction task, feature- and kernel-based approaches have been commonly used \cite{baumgartner2008concept,bui2011hybrid}. In an attempt to capture syntactic and semantic information of sentences, Murugesan, Abdulkadhar, and Natarajan~\cite{murugesan2017distributed} developed a Distributed Smoothed Tree kernel (DSTK) composed of distributed lexical parse trees and semantic feature vectors and demonstrated that the shallow linguistic information helped enhance the PPI extraction capability with the model evaluation on the five benchmark PPI corpora. 

With the recent success of deep learning in a number of applications, deep neural network models have emerged to tackle the PPI extraction task. Peng and Lu \cite{peng2017deep} have demonstrated their multichannel dependency-based convolutional neural network model (McDepCNN) effectively captures syntactic features of sentences by adding a separate channel for the dependency information of the sentence syntactic structure on the PPI task using AIMed and Bioinfer corpora. Attention mechanisms in natural language processing (NLP) have shed some light on solving long dependency issues between tokens in sequential data. The self-attention-based Transformer architecture \cite{vaswani2017attention} has proven to well preserve long-term dependencies and establish effective contextual representations. NLP models built upon Transformer architecture, such as BERT \cite{kenton2019bert}, have achieved state-of-the-art (SOTA) results in various NLP tasks, including in biology domains \cite{vig2020bertology}. Warikoo, Chang, and Hsu \cite{warikoo2021lbert} have proposed a Lexically aware BERT model (LBERT) that generates syntactic contexts emphasized representations for sentence-level bio-entity relation extraction tasks taking \textit{n}-gram parts-of-speech frames as an additional input embedding to deliver latent lexical properties, and the model outperformed the prior models on a PPI task with the five benchmark PPI corpora. Recently, Tang et al. \cite{tang2022protein} have built a PPI extraction model based on a domain-specifically pre-trained BERT and adversarial training, which showed significant improvement on the classification of the five benchmark PPI corpora.

\section{Additional PPI curation}

This section details the further curation and enhancement of the aforementioned datasets.

\subsection{Problems discovered during curation}
\label{datase-problems}

In vetting the five benchmark PPI training corpora, we identified the following problems:

\subsubsection{Bias due to restricted biological focus for each set} 
In particular, the AIMed and IEPA corpora are focused on human medical biochemistry and phenomena, including viral pathogens, whereas the set LLL is limited to a single bacterial species, {\it Bacillus subtilis}.
These differences manifest in skew and distribution of protein/gene name frequency counts between the five sets, as well as other domain-specific terminology.
In fact, the most frequently occurring protein in IEPA, {\sl insulin\/}, accounts for 14\% of the protein mentions in all of the IEPA positives, yet it does not occur in the AIMed positives set, where the most common protein, \textit{p53}, accounts for only 1.75\% of the protein names.
These sets all sampled especially different populations in the literature. 
Combining all sets together helps to counter this bias, but, in the future, we plan to collect more training data to better address this issue.

\subsubsection{Differences in notion of the definition of an \textit{interaction}}
The five sets largely restrict PPI-positive cases to clear statements of direct interaction between the two subjects. LLL further restricts positive PPI declarations to cases where a protein binds to DNA and causes or inhibits the transcription of the gene of another protein, or a statement of gene regulation---a markedly particular type of interaction.

We intentionally broaden our acceptance of a positive PPI indication. Our goal is to provide biologists with a tool to identify possible interactive connections between proteins directly from the scientific literature text.    
Because of the likelihood that claims of direct PPI will end up in future databases (if not there already), a less restrictive interpretation will allow a text mining system to report results of value that will not necessarily be found in a PPI database.

Along these lines, we did not distinguish between gene or protein for this work.
In addition to direct binding between two proteins or a protein and itself (i.e., \textit{dimers} and \textit{multimers}), we also consider interacting cases where two proteins bound to a larger complex of other proteins without necessarily contacting each other directly.

The following details an example (from the BioCreative corpus) where a direct connection between proteins \textit{PVA12} and \textit{ORP3a} is made but is not declared an actual interaction.

\begin{tcolorbox}
{\sl The targeting of the oxysterol-binding protein \textbf{ORP3a} to the endoplasmic reticulum relies on the plant VAP33 homolog \textbf{PVA12}.}
\end{tcolorbox}

On the other hand, we are mindful of the possibility of being too broad, which would result in too many PPI calls to be meaningful.

\subsubsection{Confusion over PPI-negative annotations}
\label{negative-problems}
This expanded threshold for PPI-positive impacts the public negative annotations. The following are two example cases (from AIMed corpus) where we disagree with the given negative labels.
      
\begin{tcolorbox}
{\sl In addition to this unique pathway, \textbf{FGFR3} also links to \textbf{GRB2}.}
\end{tcolorbox}

A negative interaction between proteins \textit{FGFR3} and \textit{GRB2} was declared in the public set.
  
\begin{tcolorbox}
{\sl After a brief historical incursion regarding renal artery stenosis (RAS) of renal origin, we present the main extrarenal \textbf{angiotensin}-forming enzymes, starting with \textbf{isorenin}, \textbf{tonin}, and D and \textbf{G cathepsin} and ending with the conversion enzyme and \textbf{chymase}.}
\end{tcolorbox}

In this case, negative interactions are annotated between
\textit{angiotensin} and each of \textit{isorenin}, \textit{tonin}, \textit{G cathepsin}, and \textit{chymase}, respectively, even though they are declared as forming angiotensin.

The following shows an example of a negative PPI sentence where we agree with the given label and have included in our curated set (from AIMed corpus).

\begin{tcolorbox}
{\sl The molar ratio of serum \textbf{retinol-binding protein (RBP)} to \textbf{transthyretin (TTR)} is not useful to assess vitamin A status during infection in hospitalized children.}
\end{tcolorbox}

To reduce confusion in our initial models regarding updated positive and negative relabels, we consider only those negatively labeled sentences where no positive pairs were declared in a sentence. Then, we manually examine each case to make sure we agree, disregarding (for now) those where we differ.  
For the same reason, in this work, we also disregard negative pair cases in sentences with both positive and negative annotations.
 
\subsection{Interaction Type Annotation}
PPIs aid with biological engineering. Notably, structure and protein subunit complex knowledge is critical to protein engineering, and transient interactions (e.g., chaperone to client protein) knowledge is needed for engineering at a broader scale. To make the public PPI corpora more useful for this purpose, we have added interaction type labels for the positively defined pairs in the unified datasets and the BioCreative set.
In determining the interaction type labels, we first considered top-level protein function categories from IntAct's molecular interaction ontology but discovered we lacked enough training examples to provide sufficient statistics in each of the 28 categories to properly train a model (not all interaction types occur with equal frequency).
We then tried to reduce the number of categories by making them coarser, first lowering to roughly 10 then three types. However, we found that making assignments in this manner proves too complicated with only questionable scientific value.

We finally decided on a simple binary classification with interactions being declared either {\sl enzyme\/} or {\sl structural\/} for our first pass because {\sl enzyme\/} or {\sl structural\/} accurately delineates the functional role of almost all proteins and consequently provides a concise but meaningful protein classification.
The {\sl structural\/} label is applied to protein assemblages of large, permanent cellular components, such as cell walls, histones, golgi apparatus, microtubules, membranes, and inter-cellular structures.
All other interactions are classified as \textit{enzyme}.
Type is determined by examining the given function for each protein/gene, where it can be obtained from any of several online protein databases, such as Uniprot, NCBI, and GeneCards, and from the sentence context itself.
For the five sentence-based datasets, interaction type labels are applied for positively identified protein pairs.
An example of a structural interaction label for the proteins \textit{alpha-syntrophin} and \textit{utrophin} (from BioInfer corpus) follows:

\begin{tcolorbox}
{\sl Absence of \textbf{alpha-syntrophin} leads to structurally aberrant neuromuscular synapses deficient in \textbf{utrophin}.}
\end{tcolorbox}

The remaining non-structural interactions are considered {\sl enzymatic}, a label applied to nominal enzyme activity (proteins that catalyze chemical reactions of metabolites in reaction pathways) and proteins that activate other proteins ({\sl kinases}). In this work, we also applied said label to all proteins that activated, inhibited, signaled, and formed temporary complexes with other proteins, as well as those that bind to DNA to regulate gene expression, chaperones which help proteins fold, and those that destroy proteins (proteases). The following is an example of an enzyme-labeled PPI between \textit{JAK2} and \textit{Ref-1} (from AIMed corpus):

\begin{tcolorbox}
{\sl The cytokine-activated tyrosine kinase \textbf{JAK2} activates \textbf{Raf-1} in a p21ras-dependent manner.}
\end{tcolorbox}

This process of adding type labels proved to be the most difficult and labor-intensive aspect of the training data curation with thousands of gene names and symbols that required external lookups in addition to an equally large host of specialized biological jargon and acronyms (chemical names, cell lines, experimental conditions, etc.) that required research to differentiate from proteins and establish the context necessary for understanding each sentence. Importantly, because this annotation effort is informed by resources and knowledge external to the text in question, it encodes specialized domain knowledge that makes the PPI type classification task more challenging, increasing pressure on ML models to capture sufficiently informative context adequately to make a class determination.

Appendix~\ref{sec:appendix-annot-process} shows the annotation process. Two domain experts have performed the PPI annotation and reached a high inter-annotator agreement as seen in Appendix~\ref{sec:appendix-inter-annot-agreement}. The definition of an interaction and the annotation rules were carefully determined ahead of time, according to domain expertise. Some of the rules are shown in Appendix~\ref{sec:appendix-annot-rule}, and the complete rules can be found in our GitHub repository.

\section{Methodology}

We have adopted a Transformer-based approach for the PPI classification task. In particular, we improve a relation representation exploiting the relational context information of an entity pair.

\subsection{Relation Representation augmented with Attention-based Context Information}

In a relation classification task, the \texttt{[CLS]} token is frequently used to represent a relation representation, which is a special classification token in BERT employed to capture the overall information of an input sequence. Another popular method is the entity mention pooling approach that concatenates a pair of two max-pooled entity embeddings in the last hidden state of BERT. To explicitly indicate target tokens for a relation, entity markers can be used in input, which are additional special input tokens indicating which tokens need focus for relation learning. Soares, Fitzgerald, Ling, and Kwiatkowski \cite{soares2019matching} have conducted the comparative study between marker-free and marker-embed representations showing the marker embedded approach outperforms marker-free representations on several supervised relation extraction tasks. Specifically, the concatenation of the entity start markers achieves the best performance. 

We additionally improve the relation representation built upon a pair of entities or entity start markers by adding relational context information of entities. The rationale is that additional tokens for relational context can serve a crucial role in determining the relation of the entities. For instance, the word \textit{activates} in \textit{``A activates B''} and \textit{Interaction} in \textit{``Interaction between A and B''} are important clues for the effector-effectee relation. To find the most relevant tokens for relation information, we leverage entity tokens' attention probabilities generated in the last hidden layer in BERT. We sum two entities' attention probabilities and retrieve additional tokens by the probability scores. The retrieved tokens are max-pooled then added to the final relation representation. 
\begin{gather*}
e^{attn} = max(\sum_{h}^{H}P_{attn}(e_h)) \\
rc(e_1^{attn}, e_2^{attn}) = max(\sum_{i}^{N}e_{1,i}^{attn} + e_{2,i}^{attn}) \\
\mathbf{x^r} = e^r_1 \oplus rc(e_1^{attn}, e_2^{attn}) \oplus e^r_2, 
\end{gather*}
where $P_{attn}$ denotes attention probabilities of a token. $H$ is the number of heads in the model. $rc$ stands for relation context, and $N$ is the number of tokens to be attentive. $N$ also is a hyper-parameter and is set prior to model training. In this study, $N$ is set as 20\% of an input length, which was empirically determined using validation sets of biomedical relation extraction benchmarks (see Appendix~\ref{sec:appendix_rel_ctx_eval}). $\mathbf{x^r}$ is the final relation representation for a classifier, which is the linking of entity embeddings ($e^r_1,e^r_2$) (mention pooling or entity start marker) and a max-pooled relation context embedding. When selecting tokens for relation context, we only account for alphanumerical tokens and exclude entity tokens and special tokens (besides entity markers). If a token is a part of a word (tokens with ``\#\#''), the entire word is included. Figure~\ref{model-architecture} illustrates the construction of a relation representation for a sentence with entity start markers, and the mention pooling approach is depicted in Appendix~\ref{sec:appendix_rel_rep_mention_pooling}.

\begin{figure*}
  \includegraphics[width=18cm, height=5.8cm]{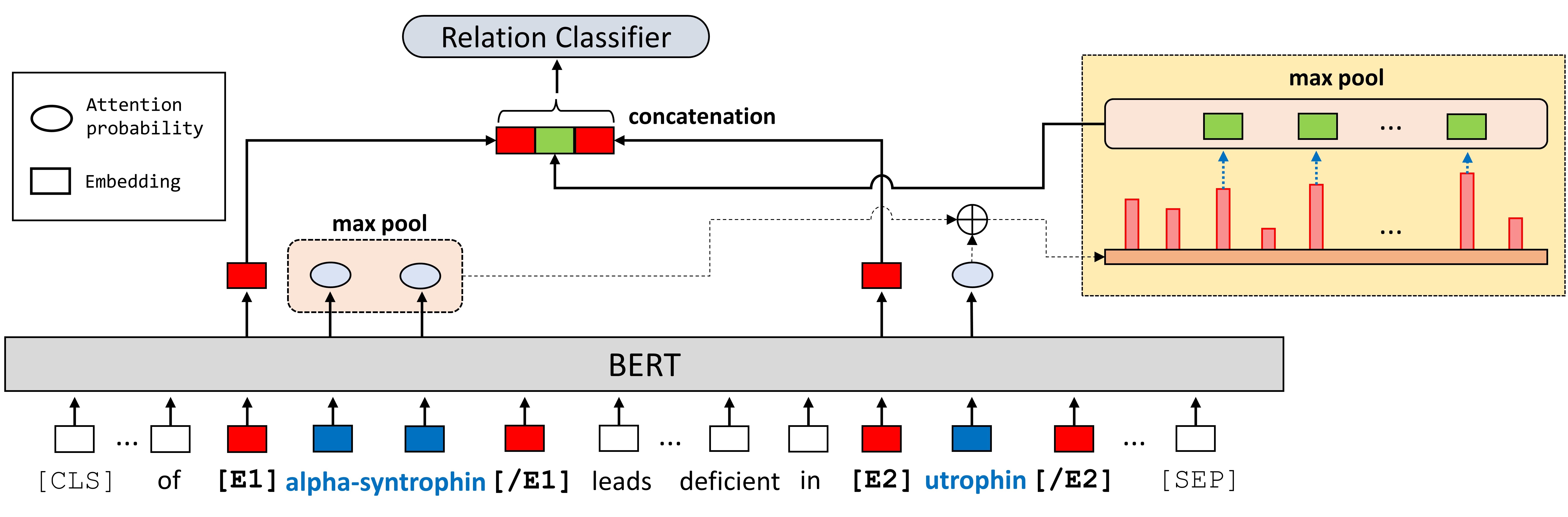}
  \centering
  \caption{The relation representation consists of entity start markers and the max-pooled of relational context, which is a series of tokens chosen by attention probability of the entities. The relation representation based on mention pooling is depicted in Appendix~\ref{sec:appendix_rel_rep_mention_pooling}. $\oplus$ denotes element-wise addition. The example sentence is \textit{Absence of alpha-syntrophin leads to structurally aberrant neuromuscular synapses deficient in utrophin}. (Source: BioInfer corpus). \label{model-architecture}}
\end{figure*}

\subsection{Model Architecture}

Our Transformer-based relation extraction model performs a sequence classification task using a logistic regression with softmax to determine the probability of relation class (e.g., $c \in \{$\textit{enzyme}, \textit{structural}, \textit{negative}$\}$) as follows:
\begin{equation}\label{first_eqn}
   P(c|X) = softmax(W\mathbf{x^r}),
\end{equation}
where $X$ and $\mathbf{x^r}$ denote examples and relation representations, respectively. The model parameters are optimized using a categorical cross entropy.
\begin{equation}\label{second_eqn}
   -\sum_{c}\delta(X,c)\log{P(c|X)},
\end{equation}
where $\delta(X,c)$ indicates whether the class of $X$ is correctly predicted ($\delta(X,c) = 1$) or not ($= 0$). Algorithm 1 illustrates the model training procedure.

\begin{algorithm}[h]
\caption{Training a PPI model}
\label{alg:algorithm}
\textbf{Initialize}: Load a pre-trained BERT model and set the max epoch and mini-batch size.\\
\textbf{Output}: Refined BERT model for PPI classification task using an attention-based relation representation. \\
\begin{algorithmic}[1] 
\STATE Given relation extraction samples, define entity spans and add entity tags when using markers.
\FOR{s in $S_{relation}$}
\STATE $D \leftarrow define\_entity\_span\_and\_add\_marker(s)$\\
\ENDFOR
\WHILE{$epoch$ \textbf{to} $epoch_{max}$}
\STATE // $b$ is a mini-batch.
\FOR{$b$ \textbf{in} D}
\FOR{\textbf{each} (\textit{e1: entity 1, e2: entity 2}) $\in$ $b$}
\STATE Generate attention-based relation representations.
\STATE $R \leftarrow e1\_emb \oplus relation\_context \oplus e2\_emb$
\ENDFOR
\STATE Produce logits.
\STATE $logits = relation\_classifier(R)$
\STATE Compute loss.
\STATE $\mathcal{L} = CrossEntropyLoss(logits, labels)$
\STATE Compute gradient and update parameters.
\STATE $\theta = \theta - \eta\nabla\theta$
\ENDFOR
\ENDWHILE
\end{algorithmic}
\end{algorithm}

\section{Experimental Setup}

We first demonstrate the effectiveness of the proposed approach on four well-known relation extraction benchmark datasets in the biomedical domain. Then, the method is evaluated on the five PPI benchmark corpora and our PPI corpus with interaction types by comparing the performance with SOTA models.

\subsection{Datasets}

In this study, we use four biomedical relation extraction (RE) datasets: ChemProt \cite{peng2018extracting}, DDI \cite{herrero2013ddi}, GAD \cite{bravo2015extraction}, and EU-ADR \cite{van2012eu}. There are various versions of the ChemProt, DDI, and GAD datasets. Here, we adopt the recent and widely used benchmark data, the Biomedical Language Understanding and Reasoning Benchmark (BLURB) provided by \cite{gu2021domain}. We also use the EU-ADR data in BioBERT \cite{lee2020biobert}. The ChemProt, DDI, and GAD datasets consist of a train/validation/test set, while the EU-ADR contains 10-fold sets for cross validation. In all of the data, target entities are anonymized with pre-defined tags, including \texttt{@GENE\$}, \texttt{@CHEMICAL\$}, \texttt{@DRUG\$}, and \texttt{@DISEASE\$}. In ChemProt and DDI, additional tags, \texttt{@CHEM-GENE\$} and \texttt{@DRUG-DRUG\$}, are used for overlapping entities. When entity markers are used, \texttt{@CHEM-GENE\$} and \texttt{@DRUG-DRUG\$} are surrounded by the \texttt{[E1-E2]} tag. Descriptions of each data follow, and Table~\ref{rel-ext-data} displays the number of data samples.

\begin{enumerate}
   \item ChemProt contains chemical-protein interactions extracted from 1,820 PubMed abstracts, and the task is evaluated using five high-level relation interaction classes: CPR:3 (UPREGULATOR), CPR:4 (DOWNREGULATOR), CPR:5 (AGONIST), CPR:6 (ANTAGONIST), and CPR:9 (SUBSTRATE). 
   \item DDI consists of drug-drug relations four relation classes (Advice, Effect, Mechanism, Int) based on 792 texts from DrugBank and 233 Medline abstracts.
   \item GAD (The Genetic Association Database corpus) contains a set of gene-disease binary associations, which was semi-automatically collected from PubMed abstracts.
   \item EU-ADR features a list of binary associations between drugs, diseases, genes, and proteins annotated on Medline abstracts.
\end{enumerate}

\begin{table}[H]
\caption{\label{rel-ext-data}
Statistics of biomedical relation extraction datasets. EU-ADR consists of 10-fold sets for cross validation.
}
\setlength{\tabcolsep}{10pt} 
\renewcommand{\arraystretch}{1.5} 
\centering
\begin{tabular}{lcccc}
\hline
 & \textbf{Train} & \textbf{Dev} & \textbf{Test} & Total\\
\hline
ChemProt    & 18,035    & 11,268    & 15,745 & 45,048 \\
DDI         & 25,296    & 2,496     & 5,716  & 33,508 \\
GAD         & 4,261     & 535       & 534    & 5,330 \\
EU-ADR      & \textit{NA} & \textit{NA} & \textit{NA} & 355 \\
\hline
\end{tabular}
\end{table}


\begin{table}[htbp]
\caption{\label{benchmark-corpora}
Five PPI benchmark corpora for \textit{positive} and \textit{negative} classes.
}
\setlength{\tabcolsep}{16pt} 
\renewcommand{\arraystretch}{1.2} 
\centering
\begin{tabular}{lcc}
\hline
\diagbox[height=1.6\line]{\textbf{Data}}{\textbf{Class}} & \textbf{Positive} & \textbf{Negative} \\
\hline
AIMed      & 1,000 & 4,834 \\
BioInfer   & 2,534 & 7,132 \\
HPRD50     & 163   & 270 \\
IEPA       & 335   & 482 \\
LLL        & 164   & 166 \\
\hline
TOTAL      & 4,196 & 12,884 \\
\hline
\end{tabular}
\end{table}

\begin{table}[htbp]
\caption{\label{typed-corpora}
Interaction typed PPI corpora for \textit{enzyme}, \textit{structural}, and \textit{negative} classes. $\dagger$ Annotations using the PPI data from BioCreative VI Track 4: Mining protein interactions and mutations for precision medicine (PM). The significant reduction from the original data in negative samples is explained in \ref{negative-problems}.
}
\setlength{\tabcolsep}{10pt} 
\renewcommand{\arraystretch}{1.2} 
\centering
\begin{tabular}{lccc}
\hline
\diagbox[height=1.6\line]{\textbf{Data}}{\textbf{Class}} & \textbf{Enzyme} & \textbf{Structural} & \textbf{Negative} \\
\hline
BioCreative VI$^\dagger$  & 378   & 83    & 0   \\
AIMed           & 548   & 182   & 1,371 \\
BioInfer        & 604   & 1,465 & 2,148 \\
HPRD50          & 103   & 34    & 87 \\
IEPA            & 271   & 2     & 224 \\
LLL             & 163   & 0     & 0 \\
\hline
TOTAL           & 2,067   & 1,766  &  3,830\\
\hline
\end{tabular}
\end{table}

\begingroup
\setlength{\tabcolsep}{12pt} 
\renewcommand{\arraystretch}{1.5} 
\begin{table*}
\caption{\label{re-datasets-results}
F1 scores on the test sets for ChemProt, DDI, GAD, and 10-fold CV for EU-ADR. In the datasets, target entities are anonymized with pre-defined tags (e.g., \texttt{@GENE\$}, \texttt{@CHEMICAL\$}, \texttt{@DRUG\$}). \textit{Mention} is a concatenation of the contextual embeddings of the entity mentions. \textit{Entity Start (markers)} are \texttt{[E1]} and \texttt{[E2]}. (\textbf{Bold}: best score in our method; \underline{underline}: best score in SOTA)
}
\centering
\small
\begin{tabular}{llcccc}
\hline
\multicolumn{2}{c}{} & \textbf{ChemProt} & \textbf{DDI} & \textbf{GAD} & \textbf{EU-ADR} \\
\hline
\textbf{SOTA} \\
\multicolumn{2}{l}{KeBioLM \cite{yuan2021improving}} & \underline{77.5} & 81.9 & \underline{84.3} & - \\
\multicolumn{2}{l}{PubMedBERT \cite{gu2021domain}} & 77.2 & \underline{83.6} & 84.1 & - \\
\multicolumn{2}{l}{BioBERT \cite{lee2020biobert} (PyTorch version)} & - & - & 82.4 & \underline{85.1} \\
\hline
\textbf{Ours} \\
\textit{Input} & \textit{Representation} \\
\multirow{3}{*}{Entity Anonymization} & \texttt{[CLS]} & 77.9 & 81.7 & 82.1 & 85.1 \\
& Mention & 78.8 & 80.0 & 83.0 & 84.2 \\
& Mention + Relation Context & \textbf{80.1} & 81.3 & \textbf{85.0} & \textbf{86.0} \\
\hdashline[1pt/3pt]
\multirow{3}{*}{Entity Anonymization + Markers} & \texttt{[CLS]} & 78.7 & 82.6 & 83.5 & 85.6 \\
& Entity Start & 76.5 & 80.7 & 82.6 & 85.0 \\
& Entity Start + Relation Context & 79.2 & \textbf{83.6} & 84.5 & 85.5 \\
\hline
\end{tabular}
\end{table*}
\endgroup

The five PPI benchmark corpora include AIMed \cite{bunescu2005comparative}, BioInfer \cite{pyysalo2007bioinfer}, HPRD50 \cite{fundel2007relex}, IEPA \cite{ding2001mining}, and LLL \cite{nedellec2005learning}. We adopt the unified version of PPI benchmark datasets provided by \cite{pyysalo2008comparative} that has been used in the SOTA models. In the datasets, the PPI relations are tagged with either \textit{positive} or \textit{negative}. The corpus statistics is described in Table~\ref{benchmark-corpora}. Our PPI annotations with interaction types (\textit{enzyme}, \textit{structural}, or \textit{negative}) are the expanded version of the five benchmark corpora and the BioCreative VI protein interaction dataset \cite{islamaj2019overview}. Table~\ref{typed-corpora} displays the corpora statistics. The annotation work in all corpora has been carried out in a sentence boundary as engaged in the five PPI benchmark corpora.

\subsection{Implementation details}
We use domain-specific pre-trained BERT models on biomedical literature, including BioBERT \cite{lee2020biobert} and PubMedBERT \cite{gu2021domain}, which has demonstrated excellent performance in biomedical NLP applications. We use PyTorch (version 1.10.2) and the HuggingFace's Transformers package (version 4.17.0) \cite{wolf2020transformers}, while the pre-trained models used are obtained from the HuggingFace model repository\footnote{\tt{\url{https://huggingface.co/models}}}. The model architecture and weight initialization follow the pre-trained models, and the hyper-parameters are tuned with the range: epoch number (3–20), batch size (8, 16), and learning rate (1e-5, 3e-5, 5e-5) with Adam. For objective comparisons, we endeavor to adopt the same models and hyperparameters used in the SOTA systems (if available) to reproduce the identical results with their relation representation. The hyperparameter details can be found in our GitHub repository. We use a dense layer with linear activation as a post-Transformer layer and train the model on the machine, Tesla V100-SXM2-32GB $\times$ 2.

\section{Results and Discussion}

\subsection{Evaluation on biomedical RE datasets}
We use the BioBERT large-cased model for the ChemProt, the PubMedBERT-uncased-fulltext model for DDI and GAD, and the BioBERT base-cased model for EU-ADR. We compare our model's performance with the SOTA results, including KeBioLM \cite{yuan2021improving} for ChemProt and GAD, PubMedBERT \cite{gu2021domain} for DDI, and BioBERT \cite{lee2020biobert} (Version\footnote{\tt{\url{https://github.com/dmis-lab/biobert-pytorch}}} as our model was built on PyTorch) for EU-ADR. KeBioLM and PubMedBERT use the combinations of entity mentions, and BioBERT uses the \texttt{[CLS]} token for relation classification. We measure the performance by the same metrics used in the SOTA systems. The results demonstrate that our proposed representation of the entity mention augmented with the relation context achieved SOTA results for ChemProt, GAD, EU-ADR, while the combination of entity start markers with the relation context produced comparable performance for DDI (shown in Table~\ref{re-datasets-results}). The relation context improves the predictions in all cases. Notably, its significance is clearly shown in EU-ADR, where we have replicated the result obtained in the SOTA model (\texttt{[CLS]}: 85.1 F1 score) using the same model, input (without markers), representation, and adding the relation context to the mention pooling, which produced a superior result over the \texttt{[CLS]} token. 

\begingroup
\setlength{\tabcolsep}{9pt} 
\renewcommand{\arraystretch}{1.5} 
\begin{table*}
\caption{\label{benchmark-ppi-result}
F1 scores via 10-fold CV on the PPI classification with the five benchmark PPI corpora. \textit{Mention} is a concatenation of the contextual embeddings of the entity mentions. \textit{Entity Start (markers)} are \texttt{[E1]} and \texttt{[E2]}. Our methods use the BioBERT base-cased model. (\textbf{Bold}: best score in our method; \underline{underline}: best score in SOTA)
}
\centering
\small
\begin{tabular}{llp{1.1cm}p{1.1cm}p{1.1cm}p{1.1cm}p{1.1cm}p{1cm}}

\hline
\multicolumn{2}{c}{} & \textbf{AIMed} & \textbf{BioInfer} & \textbf{HPRD50} & \textbf{IEPA} & \textbf{LLL} & \textit{Avg.}\\
\hline
\textbf{SOTA} \\
\multicolumn{2}{l}{DSTK \cite{murugesan2017distributed}} & 71.0 & 76.3 & 80.0 & 80.2 & \underline{89.2} & 79.3\\
\multicolumn{2}{l}{DeepResCNN \cite{zhang2019deep}} & 77.6 & 86.9 & 77.7 & 75.5 & 83.2 & 80.2\\
\multicolumn{2}{l}{LBERT \cite{warikoo2021lbert}} & 74.0 & 72.8 & \underline{85.5} & 83.7 & 86.0 & 80.4\\
\multicolumn{2}{l}{ADVBERT \cite{tang2022protein}} & \underline{83.9} & \underline{90.3} & 84.8 & \underline{84.9} & 88.7 & \underline{86.5}\\
\hline
\textbf{Ours} \\
\textit{Input} & \textit{Representation} \\
\multirow{3}{*}{Original} & \texttt{[CLS]} & 83.2 & 79.1 & 65.3 & 68.0 & 62.4 & 71.6 \\
& Mention & 90.6 & 88.0 & 83.4 & 85.2 & 84.9 & 86.4 \\
& Mention + Relation Context & 90.8 & 88.2 & 84.5 & 85.9 & 84.6 & 86.8 \\
\hdashline[1pt/3pt]
\multirow{3}{*}{Entity Markers} & \texttt{[CLS]} & 91.8 & 90.9 & 83.1 & 82.9 & 85.2 & 86.8 \\
& Entity Start & 91.4 & 90.9 & 87.3 & 86.4 & 88.8 & 89.0 \\
& Entity Start + Relation Context & \textbf{92.0} & \textbf{91.3} & \textbf{88.2} & \textbf{87.4} & \textbf{89.4} & \textbf{89.7} \\
\hline
\end{tabular}
\end{table*}
\endgroup

\begin{table}
\caption{\label{typed-ppi-result}
F1 scores via 10-fold CV on the typed PPI corpora. The BioBERT base-cased model is used.
}
\setlength{\tabcolsep}{8pt} 
\renewcommand{\arraystretch}{1.5} 
\centering
\small
\begin{tabular}{llc}
\hline
& \multicolumn{2}{r}{\textbf{Typed PPI}} \\
\hline
\textit{Input} & \textit{Representation} \\
\multirow{3}{*}{Original} & \texttt{[CLS]} & 84.7 \\
& Mention & 85.9 \\
& Mention + Relation Context & 86.4 \\
\hdashline[1pt/3pt]
\multirow{3}{*}{Entity Markers} & \texttt{[CLS]} & 85.9 \\
& Entity Start & 86.9 \\
& Entity Start + Relation Context & \textbf{87.8} \\
\hline
\end{tabular}
\end{table}

\subsection{Evaluation on PPI datasets}

We adopt BioBERT for the evaluation on the PPI data that achieved greater improvements on the performances in the recent PPI extraction works \cite{warikoo2021lbert, tang2022protein}. To compare the performance of the proposed approach with SOTA works, we evaluate our model using a 10-fold cross-validation (CV) manner and a micro F1 performance metric as adopted in the SOTA models. Table~\ref{benchmark-ppi-result} displays the evaluation results on the five benchmark PPI corpora, showing our models produce the best performances and outperform the SOTA models on the overall classification as described in the average F1 scores. Unlike the entity anonymized inputs, the inputs with entity markers perform better than the original inputs across all data, while using the \texttt{[CLS]} token in the original input performs the worst. This finding also has been observed in earlier works \cite{soares2019matching, gu2021domain}, implying the significance of explicit indication for target entities, such as markers or entity anonymization, with its type. The relation context constantly improves the performances, although a slight degradation occurred for the combination with entity mention in the LLL data, and the representation of entity start markers augmented with relation context achieves the best predictions.

In addition, we examine the model's ability on our PPI corpora with interaction types. In this experiment, we combine the six corpora where some datasets contain only single class or highly skewed samples so the model can be trained on more balanced data. The model evaluation also is carried out in a 10-fold CV manner, and Table~\ref{typed-ppi-result} reflects the micro F1 scores of each representation. The results demonstrate that the models yield consistent predictions with the best 87.8 F1 score compared to the previous experiments, and the representations augmented with relation context continually generate satisfactory outcomes. Through the observation of enhanced results on various relation extraction tasks, we can conclude that contextual representations that target entities are attentive and able to effectively provide additional information to determine the relations of entity pairs. 

\section{Conclusion}

In this work, we have augmented existing PPI corpora annotated with interaction types, which is expected to be beneficial for extracting more PPI information from scientific publications. We also have presented a Transformer architecture-based model for relation extraction. Specifically, we have improved a relation representation by adding relational context information based on entities' attention probabilities. Our models outperform SOTA models and offer proof about the effectiveness of additional relational context embedding on the biomedical relation extraction benchmarks and PPI corpora.

We will continue to improve our PPI annotations by resolving identified problems, including debiasing the training data. More examples are needed from across biological subject areas (plants, environmental, microbiomes, etc). Our goal is to provide a tool that works across all subfields of biology. Granularity in type classifications also needs to be increased, which will require more training data and manual annotation. Finally, statements of interaction that span two (or more) sentences also will require added attention in the future.

\bibliographystyle{IEEEtran}
\bibliography{main}

\begin{thebibliography}{10}
\providecommand{\url}[1]{#1}
\csname url@samestyle\endcsname
\providecommand{\newblock}{\relax}
\providecommand{\bibinfo}[2]{#2}
\providecommand{\BIBentrySTDinterwordspacing}{\spaceskip=0pt\relax}
\providecommand{\BIBentryALTinterwordstretchfactor}{4}
\providecommand{\BIBentryALTinterwordspacing}{\spaceskip=\fontdimen2\font plus
\BIBentryALTinterwordstretchfactor\fontdimen3\font minus \fontdimen4\font\relax}
\providecommand{\BIBforeignlanguage}[2]{{%
\expandafter\ifx\csname l@#1\endcsname\relax
\typeout{** WARNING: IEEEtran.bst: No hyphenation pattern has been}%
\typeout{** loaded for the language `#1'. Using the pattern for}%
\typeout{** the default language instead.}%
\else
\language=\csname l@#1\endcsname
\fi
#2}}
\providecommand{\BIBdecl}{\relax}
\BIBdecl

\bibitem{bruckner2009y2h}
A.~Br\"uckner, C.~Polge, N.~Lentze, N.~Auerbach, and U.~Schlattner, ``Yeast two-hybrid, a powerful tool for systems biology,'' \emph{International Journal of Molecular Sciences}, no.~10, pp. 2763--2788, 2009.

\bibitem{dunham2012affin}
W.~Dunham, M.~Mullin, and A.~Gingras, ``Affinity-purification coupled to mass spectrometry: basic principles and strategies,'' \emph{Proteomics}, vol.~12, no.~10, pp. 1576--90, 2012.

\bibitem{vaswani2017attention}
A.~Vaswani, N.~Shazeer, N.~Parmar, J.~Uszkoreit, L.~Jones, A.~N. Gomez, {\L}.~Kaiser, and I.~Polosukhin, ``Attention is all you need,'' in \emph{Advances in neural information processing systems}, 2017, pp. 5998--6008.

\bibitem{islamaj2019overview}
R.~Islamaj~Do{\u{g}}an, S.~Kim, A.~Chatr-Aryamontri, C.-H. Wei, D.~C. Comeau, R.~Antunes, S.~Matos, Q.~Chen, A.~Elangovan, N.~C. Panyam \emph{et~al.}, ``Overview of the biocreative vi precision medicine track: mining protein interactions and mutations for precision medicine,'' \emph{Database}, vol. 2019, 2019.

\bibitem{bunescu2005comparative}
R.~Bunescu, R.~Ge, R.~J. Kate, E.~M. Marcotte, R.~J. Mooney, A.~K. Ramani, and Y.~W. Wong, ``Comparative experiments on learning information extractors for proteins and their interactions,'' \emph{Artificial intelligence in medicine}, vol.~33, no.~2, pp. 139--155, 2005.

\bibitem{pyysalo2007bioinfer}
S.~Pyysalo, F.~Ginter, J.~Heimonen, J.~Bj{\"o}rne, J.~Boberg, J.~J{\"a}rvinen, and T.~Salakoski, ``Bioinfer: a corpus for information extraction in the biomedical domain,'' \emph{BMC bioinformatics}, vol.~8, no.~1, pp. 1--24, 2007.

\bibitem{fundel2007relex}
K.~Fundel, R.~K{\"u}ffner, and R.~Zimmer, ``Relex—relation extraction using dependency parse trees,'' \emph{Bioinformatics}, vol.~23, no.~3, pp. 365--371, 2007.

\bibitem{ding2001mining}
J.~Ding, D.~Berleant, D.~Nettleton, and E.~Wurtele, ``Mining medline: abstracts, sentences, or phrases?'' in \emph{Biocomputing 2002}.\hskip 1em plus 0.5em minus 0.4em\relax World Scientific, 2001, pp. 326--337.

\bibitem{nedellec2005learning}
C.~N{\'e}dellec, ``Learning language in logic-genic interaction extraction challenge,'' in \emph{4. Learning language in logic workshop (LLL05)}.\hskip 1em plus 0.5em minus 0.4em\relax ACM-Association for Computing Machinery, 2005.

\bibitem{pyysalo2008comparative}
S.~Pyysalo, A.~Airola, J.~Heimonen, J.~Bj{\"o}rne, F.~Ginter, and T.~Salakoski, ``Comparative analysis of five protein-protein interaction corpora,'' in \emph{BMC bioinformatics}, vol.~9, no.~3.\hskip 1em plus 0.5em minus 0.4em\relax BioMed Central, 2008, pp. 1--11.

\bibitem{tikk2010comprehensive}
D.~Tikk, P.~Thomas, P.~Palaga, J.~Hakenberg, and U.~Leser, ``A comprehensive benchmark of kernel methods to extract protein--protein interactions from literature,'' \emph{PLoS Comput Biol}, vol.~6, no.~7, p. e1000837, 2010.

\bibitem{bui2011hybrid}
Q.-C. Bui, S.~Katrenko, and P.~M. Sloot, ``A hybrid approach to extract protein--protein interactions,'' \emph{Bioinformatics}, vol.~27, no.~2, pp. 259--265, 2011.

\bibitem{warikoo2021lbert}
N.~Warikoo, Y.-C. Chang, and W.-L. Hsu, ``Lbert: Lexically aware transformer-based bidirectional encoder representation model for learning universal bio-entity relations,'' \emph{Bioinformatics}, vol.~37, no.~3, pp. 404--412, 2021.

\bibitem{baumgartner2008concept}
W.~A. Baumgartner, Z.~Lu, H.~L. Johnson, J.~G. Caporaso, J.~Paquette, A.~Lindemann, E.~K. White, O.~Medvedeva, K.~B. Cohen, and L.~Hunter, ``Concept recognition for extracting protein interaction relations from biomedical text,'' \emph{Genome biology}, vol.~9, no.~2, pp. 1--15, 2008.

\bibitem{murugesan2017distributed}
G.~Murugesan, S.~Abdulkadhar, and J.~Natarajan, ``Distributed smoothed tree kernel for protein-protein interaction extraction from the biomedical literature,'' \emph{PLoS One}, vol.~12, no.~11, p. e0187379, 2017.

\bibitem{peng2017deep}
Y.~Peng and Z.~Lu, ``Deep learning for extracting protein-protein interactions from biomedical literature,'' \emph{BioNLP 2017}, p.~29, 2017.

\bibitem{kenton2019bert}
J.~D. M.-W.~C. Kenton and L.~K. Toutanova, ``Bert: Pre-training of deep bidirectional transformers for language understanding,'' in \emph{Proceedings of NAACL-HLT}, 2019, pp. 4171--4186.

\bibitem{vig2020bertology}
J.~Vig, A.~Madani, L.~R. Varshney, C.~Xiong, N.~Rajani \emph{et~al.}, ``Bertology meets biology: Interpreting attention in protein language models,'' in \emph{International Conference on Learning Representations}, 2020.

\bibitem{tang2022protein}
Z.~Tang, X.~Guo, Z.~Bai, L.~Diao, S.~Lu, and L.~Li, ``A protein-protein interaction extraction approach based on large pre-trained language model and adversarial training,'' \emph{KSII Transactions on Internet and Information Systems (TIIS)}, vol.~16, no.~3, pp. 771--791, 2022.

\bibitem{soares2019matching}
L.~B. Soares, N.~Fitzgerald, J.~Ling, and T.~Kwiatkowski, ``Matching the blanks: Distributional similarity for relation learning,'' in \emph{Proceedings of the 57th Annual Meeting of the Association for Computational Linguistics}, 2019, pp. 2895--2905.

\bibitem{peng2018extracting}
Y.~Peng, A.~Rios, R.~Kavuluru, and Z.~Lu, ``Extracting chemical--protein relations with ensembles of svm and deep learning models,'' \emph{Database}, vol. 2018, 2018.

\bibitem{herrero2013ddi}
M.~Herrero-Zazo, I.~Segura-Bedmar, P.~Mart{\'\i}nez, and T.~Declerck, ``The ddi corpus: An annotated corpus with pharmacological substances and drug--drug interactions,'' \emph{Journal of biomedical informatics}, vol.~46, no.~5, pp. 914--920, 2013.

\bibitem{bravo2015extraction}
{\`A}.~Bravo, J.~Pi{\~n}ero, N.~Queralt-Rosinach, M.~Rautschka, and L.~I. Furlong, ``Extraction of relations between genes and diseases from text and large-scale data analysis: implications for translational research,'' \emph{BMC bioinformatics}, vol.~16, no.~1, pp. 1--17, 2015.

\bibitem{van2012eu}
E.~M. Van~Mulligen, A.~Fourrier-Reglat, D.~Gurwitz, M.~Molokhia, A.~Nieto, G.~Trifiro, J.~A. Kors, and L.~I. Furlong, ``The eu-adr corpus: annotated drugs, diseases, targets, and their relationships,'' \emph{Journal of biomedical informatics}, vol.~45, no.~5, pp. 879--884, 2012.

\bibitem{gu2021domain}
Y.~Gu, R.~Tinn, H.~Cheng, M.~Lucas, N.~Usuyama, X.~Liu, T.~Naumann, J.~Gao, and H.~Poon, ``Domain-specific language model pretraining for biomedical natural language processing,'' \emph{ACM Transactions on Computing for Healthcare (HEALTH)}, vol.~3, no.~1, pp. 1--23, 2021.

\bibitem{lee2020biobert}
J.~Lee, W.~Yoon, S.~Kim, D.~Kim, S.~Kim, C.~H. So, and J.~Kang, ``Biobert: a pre-trained biomedical language representation model for biomedical text mining,'' \emph{Bioinformatics}, vol.~36, no.~4, pp. 1234--1240, 2020.

\bibitem{yuan2021improving}
Z.~Yuan, Y.~Liu, C.~Tan, S.~Huang, and F.~Huang, ``Improving biomedical pretrained language models with knowledge,'' in \emph{Proceedings of the 20th Workshop on Biomedical Language Processing}, 2021, pp. 180--190.

\bibitem{wolf2020transformers}
T.~Wolf, L.~Debut, V.~Sanh, J.~Chaumond, C.~Delangue, A.~Moi, P.~Cistac, T.~Rault, R.~Louf, M.~Funtowicz \emph{et~al.}, ``Transformers: State-of-the-art natural language processing,'' in \emph{Proceedings of the 2020 conference on empirical methods in natural language processing: system demonstrations}, 2020, pp. 38--45.

\bibitem{zhang2019deep}
H.~Zhang, R.~Guan, F.~Zhou, Y.~Liang, Z.-H. Zhan, L.~Huang, and X.~Feng, ``Deep residual convolutional neural network for protein-protein interaction extraction,'' \emph{IEEE Access}, vol.~7, pp. 89\,354--89\,365, 2019.

\end{thebibliography}

\appendices

\section{Annotation process diagram}
\label{sec:appendix-annot-process}
\begin{figure}[H]
  \includegraphics[width=9cm, height=9cm]{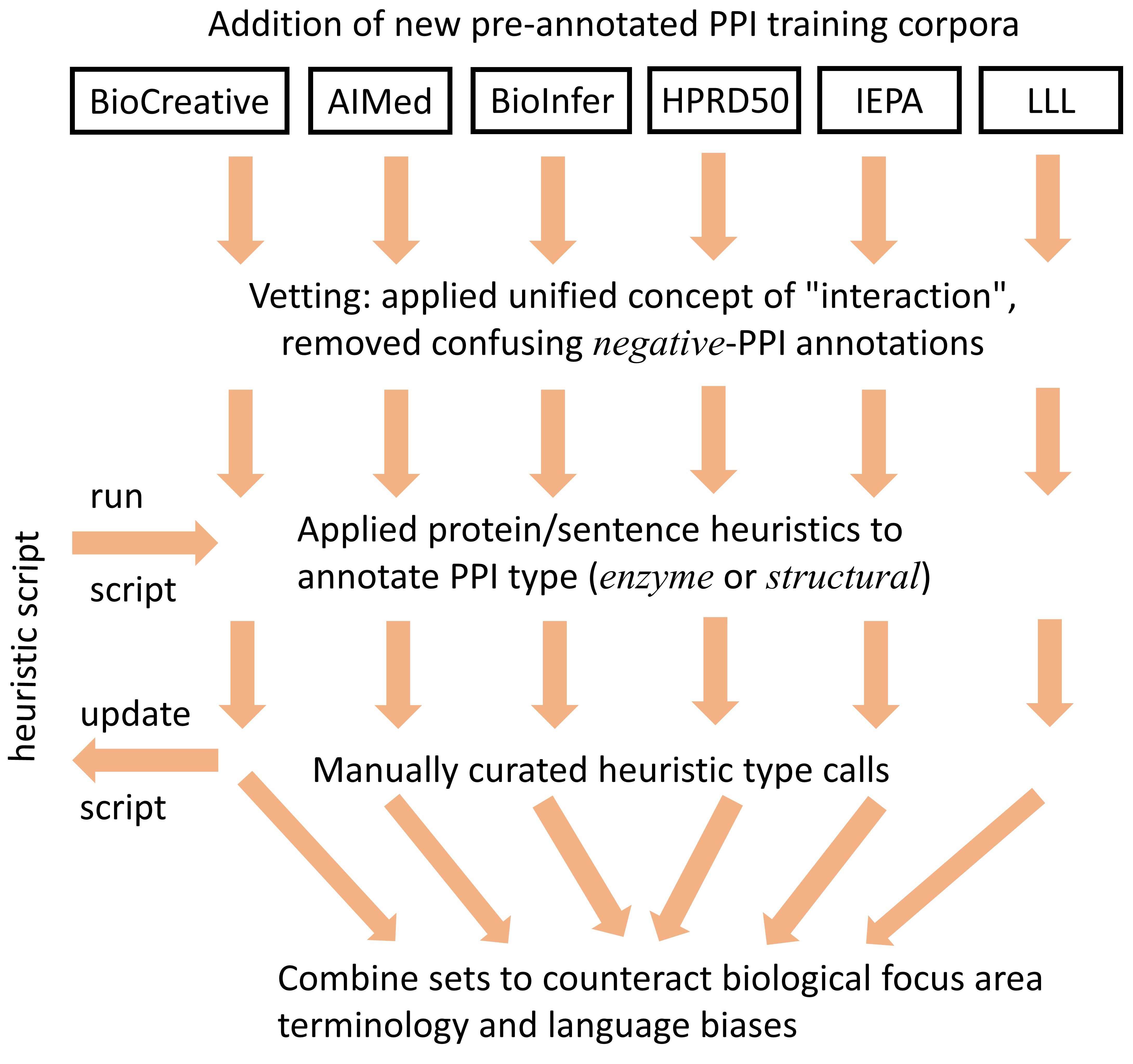}
  \centering
\end{figure}

\section{Inter-annotator agreement}
\label{sec:appendix-inter-annot-agreement}
We measured the inter-annotator agreement scores to observe the discrepancy between the annotators in the PPI relation types. The annotated data statistics can be found in Table~\ref{typed-corpora}. As seen in Table~\ref{inter-annot-scores}, the two annotators achieved a high inter-annotator agreement.

\begin{table}[h]
\caption{\label{inter-annot-scores}
Inter-annotator agreement statistics between the two annotators for the three PPI types.
}
\setlength{\tabcolsep}{15pt} 
\renewcommand{\arraystretch}{1.2} 
\centering
\begin{tabular}{lcc}
\hline
\textbf{Relation type} & \textbf{A1} & \textbf{A2} \\
\hline
\textit{enzyme} \\
A1 & \textit{NA} & 0.92 \\
A2 & 0.92 & \textit{NA} \\
\hdashline[1pt/3pt]
\textit{structural} \\
A1 & \textit{NA} & 0.90 \\
A2 & 0.90 & \textit{NA} \\
\hdashline[1pt/3pt]
\textit{negative} \\
A1 & \textit{NA} & 0.95 \\
A2 & 0.95 & \textit{NA} \\
\hline
\end{tabular}
\end{table}

\begingroup
\setlength{\tabcolsep}{8pt} 
\renewcommand{\arraystretch}{1.5} 
\begin{table*}
\caption{\label{rel-ctx-size-results}
F1 scores on the validation set for ChemProt, DDI, GAD, and EU-ADR with different sizes of relation context: 10\%, 20\%, and 30\% of an input length (except for tokens to be ignored).
}
\centering
\small
\begin{tabular}{lccccc}
\hline
\multirow{2}{*}{} & \textbf{ChemProt} & \textbf{DDI} & \textbf{GAD} & \textbf{EU-ADR} & \textit{Avg.} \\
\cline{2-6}
& 10\%/20\%/30\% & 10\%/20\%/30\% & 10\%/20\%/30\% & 10\%/20\%/30\% & 10\%/20\%/30\% \\
\hline
Mention + Relation Context & 82.2/\textbf{82.3}/81.7 & 85.1/\textbf{87}/85.1 & 83.9/\textbf{84.6}/84.2 & \textbf{86.3}/86.2/85.8 & 84.4/\textbf{85.0}/84.2\\
\hdashline[1pt/3pt]
Entity Start + Relation Context & 81.8/\textbf{83.4}/82.9 & 86.6/\textbf{86.8}/83.4 & 83.7/\textbf{84.4}/\textbf{84.4} & 88.5/\textbf{88.9}/88.6 & 85.2/\textbf{85.9}/84.8 \\
\hline
\end{tabular}
\end{table*}
\endgroup

\begin{figure*}
  \includegraphics[width=18cm, height=7cm]{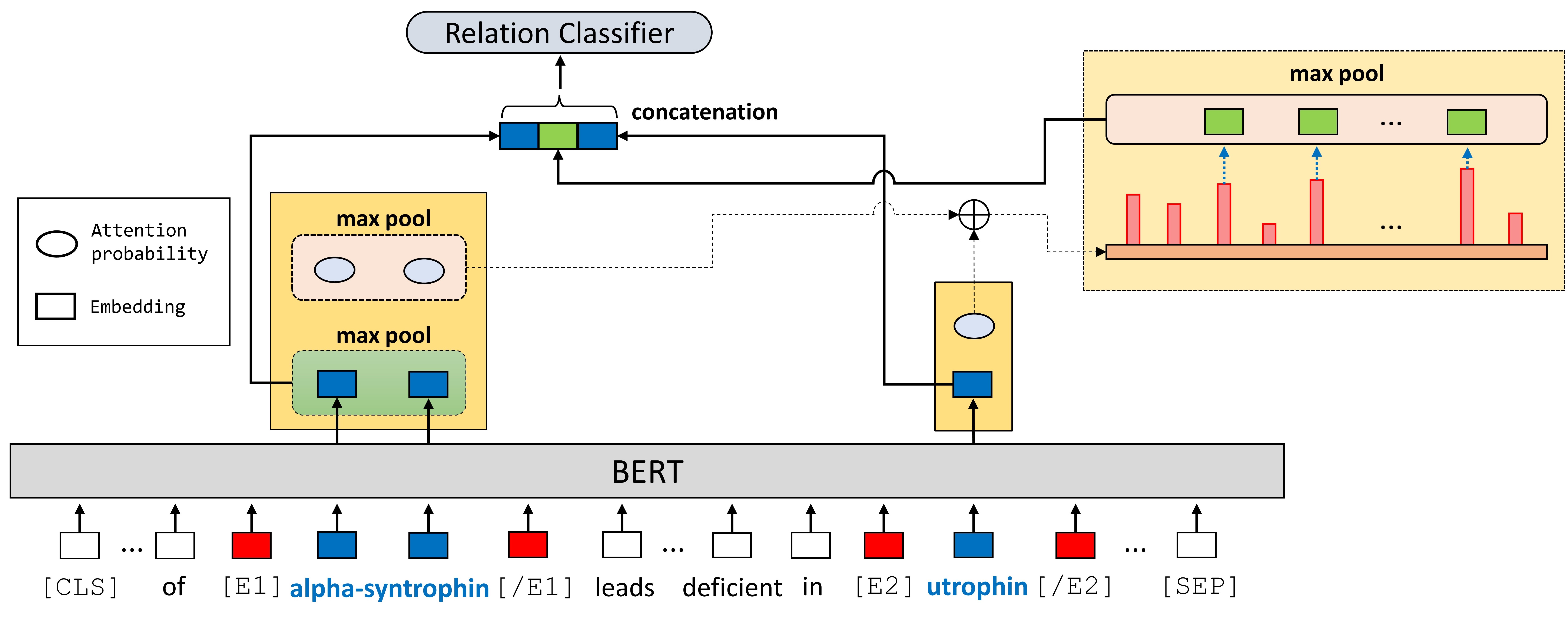}
  \centering
  \caption{The relation representation consists of the max-pooled of two entity contextualized embeddings and the max-pooled of relational context, which is a series of tokens chosen by attention probability of the entities. $\oplus$ denotes element-wise addition. The example sentence is \textit{Absence of alpha-syntrophin leads to structurally aberrant neuromuscular synapses deficient in utrophin}. (Source: BioInfer corpus). \label{model-architecture-mention-pooling}}
\end{figure*}

\section{Annotation rule examples}
\label{sec:appendix-annot-rule}

\begin{enumerate}
   \item Proteins/Genes ending in –in or –ins are pre-identified as structural (actin, catenin, …).
    Exceptions include:
    \begin{enumerate}
        \item Toxin
        \item Beta-catenin (can be gene regulator OR structural as it is a dual-function gene)
        \item Calreticulin – multifunction; mostly enzyme.
    \end{enumerate}
   \item Histones and nucleosomes are not considered structural because their ``structure'' is mutable and controls regulation.
   \item Proteins/Genes ending in –ase are preidentified as enzymes.
   \item Proteins/Genes containing inhibitor, activator, transcription factor, repressor, enhancer, or regulator are preidentified as enzymes.
\end{enumerate}

\section{Evaluation on different relation context sizes}
\label{sec:appendix_rel_ctx_eval}

To find an appropriate size of attentive context of target entities, we evaluated different sizes of relation context using the biomedical relation extraction benchmark datasets: ChemProt, DDI, GAD, and EU-ADR. We leveraged 10\%, 20\%, and 30\% of a sequence length for a number of attentive tokens of target entities and compared them on the respective validation set of the datasets. When selecting tokens for relation context, we only account for the alphanumerical tokens and exclude entity tokens (e.g., \texttt{[CLS]}; \texttt{[SEP]}) and special tokens (besides entity markers). Because the EU-ADR is a 10-fold cross validation set, we split a training set in each fold in a 9:1 ratio, i.e., 90\% of the data are used for training the model, while 10\% are used for validating the model. Without using a test set, the average scores of cross validations on train/validation sets were measured. Table~\ref{rel-ctx-size-results} demonstrates the F1 scores of different sizes of relation context, and 20\% of an input length---except for tokens to be ignored---showed the best performances on both entity mention use and entity start marker use in representation.

\section{Relation representation using mention pooling}
\label{sec:appendix_rel_rep_mention_pooling}

Figure~\ref{model-architecture-mention-pooling} illustrates the construction of a relation representation for a sentence using mention pooling. As in the entity start marker method, input sentences are tagged with entity markers. The rectangles and ovals represent the tokens' embeddings and attention probabilities, respectively.

\end{document}